\documentclass[12pt,preprint]{aastex}

\newcommand{\FeII}{[\ion{Fe}{2}]}
\newcommand{\SII}{[\ion{S}{2}]}
\newcommand{\OI}{[\ion{O}{1}]}
\newcommand{\NII}{[\ion{N}{2}]}
\newcommand{\kms}{~{\rm km~s^{-1}}}
\newcommand{\dotsec}{\rlap.^{''}}
\newcommand{\degree}{^{\circ}}

\slugcomment{}

\shorttitle{\FeII~$\lambda$1.644 $\micron$ SPECTROSCOPY OF HL TAU}
\shortauthors{Pyo et al.}

\singlespace
\begin{document}

\title{ADAPTIVE OPTICS SPECTROSCOPY OF THE \FeII\ OUTFLOWS FROM HL TAURI AND RW AURIGAE\footnote{Based on data collected at the Subaru Telescope, which is operated by the National Astronomical Observatory of Japan.}}

\author{T{\sc ae}-S{\sc oo} P{\sc yo}\altaffilmark{2}, M{\sc asahiko} H{\sc ayashi}\altaffilmark{2,3}, N{\sc aoto} K{\sc obayashi}\altaffilmark{4}, A{\sc lan} T. T{\sc okunaga}\altaffilmark{5}, H{\sc iroshi} T{\sc erada}\altaffilmark{2}, H{\sc ideki} T{\sc akami}\altaffilmark{2,3}, N{\sc aruhisa} T{\sc akato}\altaffilmark{2,3}, C{\sc hristoper} J. D{\sc avis}\altaffilmark{6}, M{\sc ichihiro} T{\sc akami}\altaffilmark{2}, S{\sc aeko} S. H{\sc ayashi}\altaffilmark{2,3}, W{\sc olfgang} G{\sc \"{a}ssler}\altaffilmark{2,7}, S{\sc hin} O{\sc ya}\altaffilmark{2}, Y{\sc utaka} H{\sc ayano}\altaffilmark{8}, Y{\sc ukiko} K{\sc amata}\altaffilmark{8}, Y{\sc osuke} M{\sc inowa}\altaffilmark{9},M{\sc asanori} I{\sc ye}\altaffilmark{3,8}, T{\sc omonori} U{\sc suda}\altaffilmark{2,3}, T{\sc akayuki} N{\sc ishikawa}\altaffilmark{3}, and K{\sc o} N{\sc edachi}\altaffilmark{9}}

\altaffiltext{2}{Subaru Telescope, National Astronomical Observatory of Japan, 650 North A`oh\=ok\=u Place, Hilo, HI 96720, USA} 
\altaffiltext{3}{School of Mathematical and Physical Science, The Graduate University for Advanced Studies (SOKENDAI), Hayama, Kanagawa 240-0193, Japan}
\altaffiltext{4}{Institute of Astronomy, University of Tokyo, Mitaka, Tokyo 181-0015, Japan} 
\altaffiltext{5}{Institute for Astronomy, University of Hawaii, 2680 Woodlawn Drive, Honolulu, HI 96822, USA}
\altaffiltext{6}{Joint Astronomy Centre, University Park, 660 North A`oh\=ok\=u Place, Hilo, HI 96720, USA}
\altaffiltext{7}{Max-Planck-Institut f\"{u}r Astronomie, K\"{o}nigstuhl 17, Heidelberg D-69117, Germany}
\altaffiltext{8}{National Astronomical Observatory of Japan, Mitaka, Tokyo 181-8588, Japan}
\altaffiltext{9}{Department of Astronomy, School of Science, University of Tokyo, Tokyo 113-0033, Japan}

\email{pyo@subaru.naoj.org}

\begin{abstract}
We present new results of \FeII\ $\lambda$1.644~$\micron$ spectroscopy toward the jets from HL~Tau and RW~Aur carried out with the Subaru Telescope combined with the adaptive optics system.
We observed the regions within 2$^{''}$--3$^{''}$ from the stars with the sub-arcsecond resolutions of 0$\rlap.^{''}$5 and 0$\rlap.^{''}$2 for HL~Tau and RW~Aur, respectively.
In addition to the strong, high velocity emission extended along each jet, we detected a blueshifted low velocity emission feature seen as a wing or shoulder of the high velocity emission at each stellar position.
Detailed analysis shows that the low velocity emission of HL~Tau is peaked at 0$\dotsec$077 $\pm$ 0$\dotsec$04 away from the stellar position, or 15~AU of actual distance from the star.
We did not confirm such displacement for RW~Aur in its position velocity diagram (PVD).
The PVDs of HL~Tau and RW~Aur show a characteristic similar to those of the cold disk wind and X-wind models in that the [\ion{Fe}{2}] line width is broad in the vicinity of the stellar position and is narrower at the extended jet.
A closer comparison suggests, however, that the disk wind model tends to have too large line width at the jet while the X-wind model has excess emission on the redshifted side at the stellar position.
The narrow velocity width with symmetric line profiles of the observed high velocity emission supports an X-wind type model where the launching region is localized in a small radial range, while the low velocity emission located away from the star favors the presence of a disk wind.  
The \FeII\ emission from the HL~Tau jet shows a gap of 0$\dotsec$8 between the redshifted jet and the star, indicating the presence of an optically thick disk of $\sim$160 AU in radius. 
The \FeII\ emission from the RW~Aur jet shows a marked drop from the redshifted peak at $Y \sim-$0$\dotsec$2 toward the star, suggesting that its disk radius is smaller than 40~AU. 
The normalized Br12 emission of HL Tau shows a significantly large deviation from the normalized continuum in spatial profile along the blueshifted jet. 
It means that part of the Br12 emission originates in the blueshifted jet itself.  
\end{abstract}

\keywords{ISM: Herbig-Haro objects --- ISM: individual (HL Tauri:HH 150, RW Aurigae:HH 229) --- ISM: jets and outflows --- stars: formation --- stars: pre-main sequence --- techniques: high angular resolution}

\section{INTRODUCTION}
Spectroscopy with high angular and spectral resolutions is essential for understanding the driving mechanism of the outflows from young stellar objects.
In previous contributions, we demonstrated that such observations with the near-infrared \FeII\ $\lambda$1.644~$\micron$ line is a powerful tool to study the acceleration and collimation regions \citep{pyo02, pyo03, pyo05}.
In the position-velocity diagrams (PVDs) of L1551 IRS~5 and DG~Tau toward their blueshifted jets, the blueshifted \FeII\ emission showed two clearly distinct components in velocity and space: the high velocity component (HVC: $V=$~200--400~$\kms$) is extended and has a narrow line width, while the low velocity component (LVC: $V=$~50--200~$\kms$) is observed closer to the driving sources and shows a larger line width.
Because of the distinct natures of these two velocity components, we suggested that they may be two physically different outflows: the HVC is a highly collimated jet, while the LVC is a widely opened disk wind.

HL Tau is a young star in a transitional phase from an embedded protostar to a exposed T~Tauri star, as has been suggested from its large amount of circumstellar material, strong outflows and optical nebulosity.
Its ``flat'' spectral energy distribution from the near-infrared to far-infrared together with its strong optical veiling suggests that HL~Tau is still in the active accretion phase.
This is also supported by the presence of an infalling envelope of $\sim$1000~AU as a remnant of its placental molecular cloud core \citep{HOM93}, although the infalling interpretation may not be unique because the flattened structure around HL~Tau overlaps with the wall of a bubble evacuated by XZ~Tau in a large scale CO map \citep{Welch00}.
Millimeter and sub-millimeter interferometry has revealed a circumstellar disk ($\sim$100~AU) inside the envelope \citep{lay97,mundy96,lay94}.
Its jet was discovered by \citet{MRB88}.
The inclination angle of the disk with respect to the plane of the sky is $\sim$42$^{\circ}\pm$5$^{\circ}$ \citep{lay97}.
{\it Hubble Space Telescope} observations revealed that the optical continuum emission from HL~Tau is not from the star itself but from a reflection nebula surrounding the blueshifted jet \citep{Sta95}. 
\citet{close97} showed in their adaptive optics observations that the near-infrared point source coincides with the star itself.

RW~Aur is a multiple star system located in the Taurus-Auriga molecular cloud.
The brightest stellar component is called ``A,'' which is the source of the outflow called HH 229.
RW~Aur~A itself may be a spectroscopic binary \citep{gahm1999,petrov01}.
The fainter component ``B'' is located 1$\dotsec$5 away from A at a position angle of 258$\degree$.
Although RW~Aur~B was also known as a close binary system, its companion ``C'' was shown to be a false source \citep{WG01}.
RW~Aur~A and B are both classical T~Tauri stars (CTTS) with active accretion suggested by their strong H$\alpha$ emission in the optical spectroscopy \citep{Duchene99} and large veiling in the near-infrared wavelengths \citep{FE99}.
The inclination of the jet axis is $\sim$46$^{\circ}\pm$3$^\circ$ degrees with respect to the line of sight \citep{LCD03}.
Transverse velocity shifts across the jet axis were detected by \citet{coffey04}, who interpreted it as jet rotation around its axis, although this rotation appears to be in the opposite sense to the rotation of the disk \citep{cabrit05}.

In this article, we present new results of the \FeII\ $\lambda$1.644~$\micron$ line spectroscopy toward the outflows from HL~Tau and RW~Aur~A.
The data were obtained with sub-arcsecond resolutions using the adaptive optics system, allowing us to study the spatial and radial velocity structure of the outflows in the vicinity of their driving stars.
We assume the distance of 140~pc to HL~Tau and RW~Aur in this paper.

\section{OBSERVATIONS AND DATA REDUCTION}
The observations were conducted on 2001 December 25 (UT) for HL~Tau and 2002 November 26 (UT) for RW~Aur using the Infrared Camera and Spectrograph \citep[IRCS]{koba00,toku98} combined with the adaptive optics (AO) system \citep{takami04}, both mounted together at the Cassegrain focus of the Subaru Telescope atop Mauna Kea, Hawaii.
The observations of the HL Tau jet were performed as part of the commissioning run of IRCS. 
The $H$-band spectra were taken using the IRCS echelle spectrograph equipped with a Raytheon 1024$\times$1024 InSb array with an Aladdin II multiplexer. 
The pixel scale along the slit was 0$\rlap.^{''}$060  pixel$^{-1}$. 

For the HL Tau jet (HH 150), the slit was 0$\rlap.^{''}$6 (W) $\times$ 3$\rlap.^{''}$8 (L) and the resulting velocity resolution was 60 $\kms$. 
The position angle of the slit was 51$^{\circ}$. 
The total on-source exposure time was 2100~s.
Sky spectra were obtained at 60$^{''}$ north of the object.
We observed the standard star HD 25175 (A0V, T$_{eff} =$ 9480 K, $V =$ 6.311 mag) for calibrating flux and removing telluric absorptions.
For the RW Aur jet, the slit was 0$\rlap.^{''}$3 (W) $\times$ 5$\rlap.^{''}$6 (L) and the resulting velocity resolution was 30 $\kms$. 
The position angle of the slit was 120$^\circ$, which was found to be 10$^\circ$ smaller than the actual position angle of the jet after the observations.
The total on-source exposure time was 960~s.
Sky spectra were obtained at 15$^{''}$ north and 20$^{''}$ east of the object.
The standard star was HD 39357 (A0V, T$_{eff} =$ 9480 K, $V =$ 4.557) for  calibration. 

We used HL Tau ($R =$ 13.53) and RW Aur ($R =$ 9.50) themselves\footnote{The $R-$magnitudes are from \citet{HP92} and \citet{elias78}.} as the reference stars of wavefront correction for the AO.
The AO correction depends on the magnitude of a reference star and the natural seeing size \citep{takami04}.
The AO system allowed us to achieve a sub-arcsecond resolution of 0$\rlap.^{''}$5 (FWHM) (Strehl ratio $\sim$ 0.01) in the $H$-band for HL Tau under a poor seeing condition exceeding 1$^{''}$ at the time of observations.
The AO corrected angular resolution was 0$\rlap.^{''}$2 (Strehl ratio $\sim$ 0.025) for RW~Aur.

We reduced the data using the IRAF packages as described in Paper~I\footnote{Details of the data reduction method for IRCS echelle spectra are available at http://www.naoj.org/staff/pyo/IRCS\_and\_Reduction/IRCS\_reduction\_html.html.}.
We subtracted the stellar continuum from the spectra. 
The subtraction was performed by the BACKGROUND task with a fourth order polynomial fitting over the range from 1.634~$\micron$ to 1.676~$\micron$, which was the coverage of the 34th order of the echelle spectrum.
We show for reference in Figure~\ref{fig1} the original spectra of HL~Tau and RW~Aur before continuum subtraction integrated over $\pm$0$\dotsec$25 around each star in comparison with the photospheric reference spectrum of HR8086. 

\section{RESULTS}

\subsection{HL Tau}

\begin{figure}
\epsscale{0.9}
\plotone{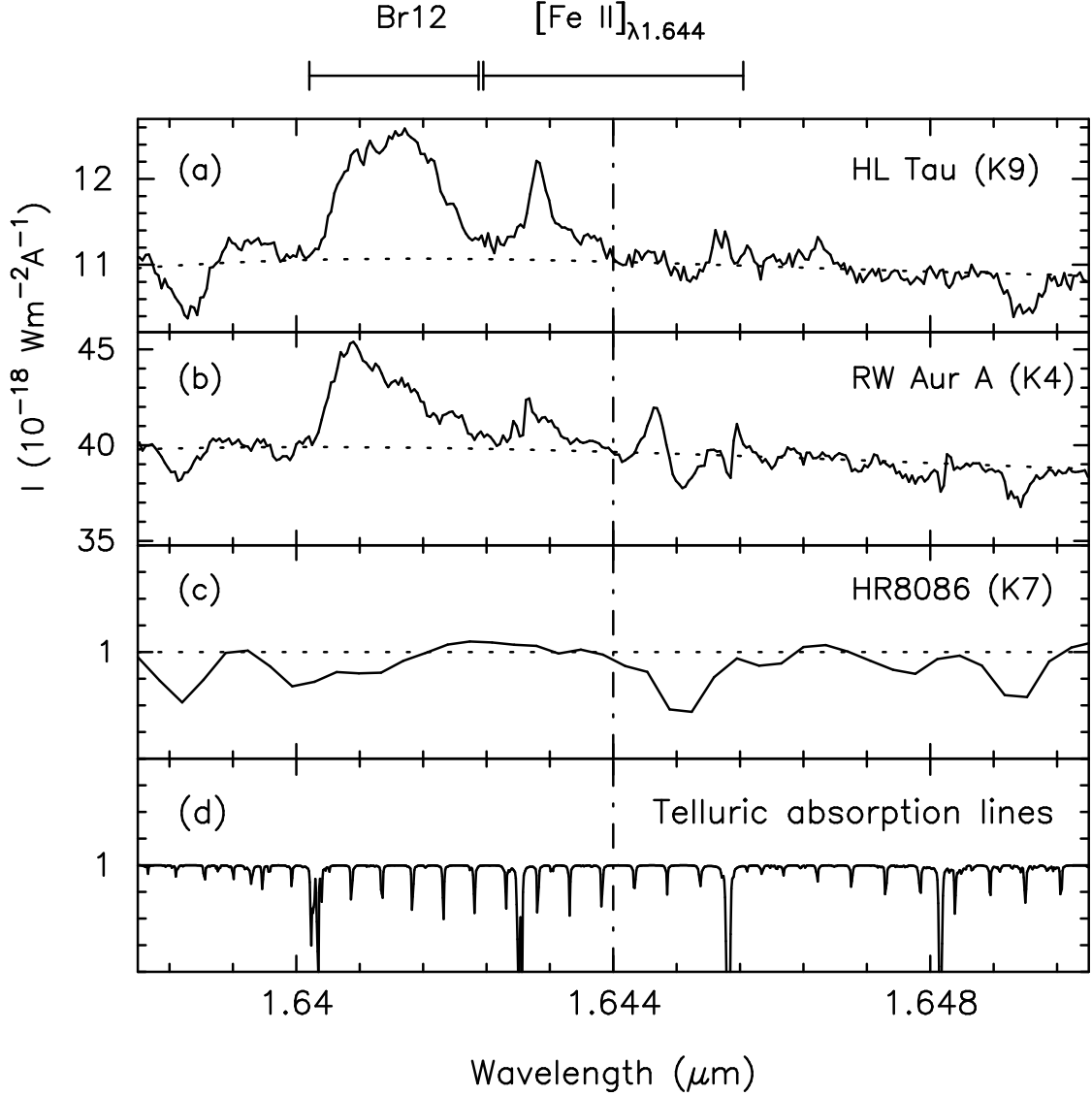}
\caption{$H$-band echelle spectra around \FeII\ $\lambda$1.644~$\mu$m of ({\it a}) HL~Tau  and ({\it b}) RW~Aur. 
Intensities are integreated over $\pm$0$\dotsec$25 around the central sources. 
Dotted lines indicate the fitted continuum levels. 
({\it c}) Normalized $H$-band spectrum of HR 8086 (K7V) with R~$=$~3000 taken from \citet{meyer98} for identification of photospheric absorption lines. 
The spectrum of HR 8086 is shifted toward the longer wavelength by $\delta \lambda =$~0.00127~$\mu$m so that its photospheric absorption line wavelengths match those of HL~Tau and RW~Aur. 
({\it d}) Telluric absorption lines.
The dash-dotted vertical line corresponds to $V_{\rm LSR}=$~0~$\kms$ for the \FeII\ $\lambda$1.644~$\mu$m line. 
Positions of the Br12, \FeII\ $\lambda$1.644~$\mu$m are marked on the upper side of the figure.
The horizontal line for the \FeII\ line shows the vlocity range of $-$300 to $+$300~$\kms$. 
\label{fig1}}
\end{figure}

\begin{figure}
\plotone{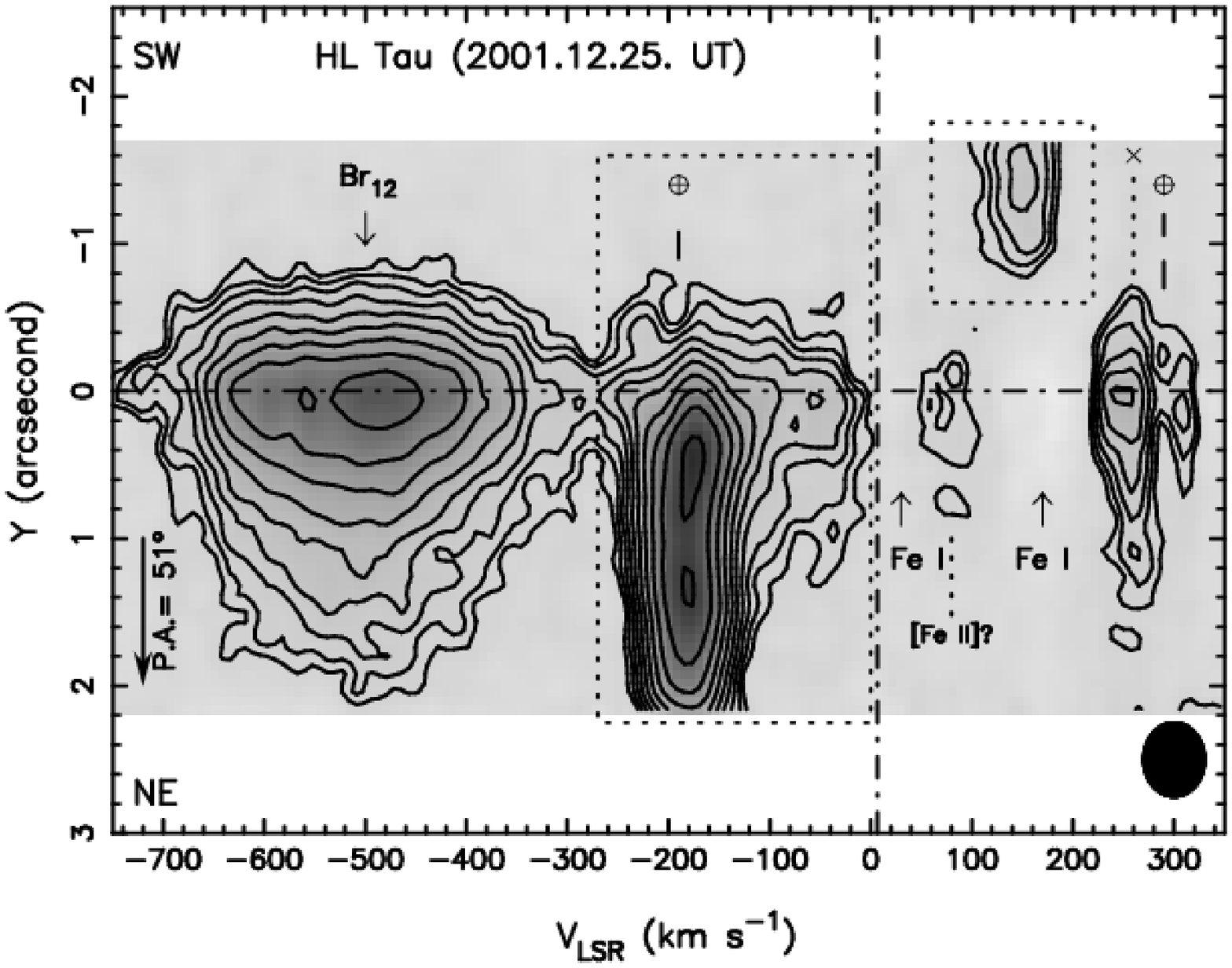}
\caption{Continuum-subtracted position velocity diagram of the \FeII\ $\lambda$1.644~$\micron$ emission toward the jet from HL~Tau.
The data was smoothed by a Gaussian with ${\sigma}_G~=$1.1 pixels. 
Contours are shown from 0.01 ($\sim$3$\sigma$) to 0.22 ($\sim$73$\sigma$) $\times$ 10$^{-18}$ W m$^{-2}$ \AA$^{-1}$ with equal intervals in a logarithmic scale.
The axis $Y$ is along the slit with a position angle of 51$^\circ$. 
The region with positive $Y$ corresponds to the northeast of HL Tau. 
The \FeII\ emission features are indicated by dotted line boxes. 
The dash-dotted vertical line indicates the systemic velocity at $V_{\rm LSR} \sim +$6.4 $\kms$ \citep{HOM93}. 
The dash-dotted horizontal line indicates the position of the continuum peak corresponding to the stellar position of HL~Tau.
The strong and extended emission at $V_{\rm LSR} < -$280 $\kms$ is the Br12 emission.
Photospheric \ion{Fe}{1} lines are indicated by arrows and telluric lines are shown by $\oplus$. 
Non-\FeII\ emission is marked by a cross ($\times$). 
The filled ellipse at the lower right shows the velocity and spatial resolutions of 60~$\kms$ and 0$\rlap.^{''}$5, respectively.
\label{fig2}}
\end{figure}

A continuum subtracted position-velocity diagram (PVD) of the \FeII\ $\lambda$1.644~$\micron$ emission is shown in Figure~\ref{fig2}, where $Y$ is the angular distance measured along the optical jet (P.A. $=$ 51$\degree$) from the $H$ band continuum peak, the stellar position of HL~Tau \citep{close97}.
The angular and velocity resolutions were 0$\dotsec$5 and 60$\kms$, respectively.
The PVD shows a blueshifted feature at $V_{\rm LSR}=-$260 to 0~$\kms$ for $Y>-0 \dotsec 8$ and a redshifted feature  at $+$100 to $+$200~$\kms$
for $Y<-0 \dotsec 8$.
We regard these blueshifted and redshifted features, enclosed by dotted line boxes in the figure, as the \FeII\ emission originating in the jet.
There could be another weak \FeII\ feature at $V_{\rm LSR}=+$80~$\kms$ around $Y=$ 0$''$, but this feature is not significant in Figure~\ref{fig1}.
It is affected by the photospheric \ion{Fe}{1} absorption lines at $+$20 and $+$170~$\kms$ with respect to the rest wavelength of \FeII\ \citep{hamann94b}.
In addition to the contamination by the absorption lines, the weak \FeII\ cannot be distinguished from other residuals of stellar continuum subtraction; many emission and photospheric absorption features are seen on the stellar continuum in the range $-0\dotsec 6 \la Y \la 0\dotsec 8$, making it difficult to determine the true continuum baseline there (see Figure~\ref{fig1}).
There is an unidentified weak emission feature at $V_{\rm LSR}=+$220 to 320~$\kms$ and $Y=-$0$\dotsec$7 to 1$\dotsec$5 superposed on a narrow telluric absorption line.
A broad Br12 line is conspicuous in addition to the \FeII\ jet emission.
Note that the $V_{\rm LSR}$ of Figure~\ref{fig2} is with respect to the \FeII\ $\lambda$1.644 $\mu$m line and not to the Br12 emission line.

At the northeast of HL Tau ($Y>0 \arcsec$), the entire \FeII\ emission is blueshifted with respected to the systemic radial velocity of $V_{\rm LSR}\sim+6.4~\kms$ \citep{HOM93}, except for the marginal feature at $V_{\rm LSR}=+$80~$\kms$.
The blueshifted \FeII\ emission is dominated by a strong, narrow feature at $V_{\rm LSR}=-$180~$\kms$ extending along the jet.
Within 1$\arcsec$ from the star, the \FeII\ emission is accompanied by a less blueshifted wing or shoulder, which has a marginal peak at $V_{\rm LSR}=-50~\kms$.
By analogy with the cases of jets from L1551 IRS~5 and DG~Tau \citep{pyo02, pyo03, pyo05}, we call the more blueshifted, stronger and extended component as the high velocity component (HVC) and the less blueshifted wing or shoulder feature as the low velocity component (LVC).
The intensity of the LVC, although seven times weaker than that of the HVC at $Y=0''$, is much larger than the baseline fluctuation caused by the uncertainty of the continuum level and is therefore real. 
Note that the LVC might be connected to the marginal feature at $V_{\rm LSR}=+$80~$\kms$ if there were no photospheric absorption by \ion{Fe}{1}.

A redshifted counter jet is seen at the southwest of HL~Tau ($Y<-$0\rlap.{$''$}8).
This emission has a peak velocity of $V_{\rm LSR}=+$150~$\kms$ with a width similarly narrow to that of the blueshifted jet.
The 0\rlap.{$''$}8 gap between the jet and star is a natural consequence if the base of the redshifted jet is occulted by a circumstellar disk.
With the assumed jet/disk inclination of 45$^\circ$, it leads to a disk radius of 160~AU, consistent with mm and sub-mm interferometric measurements \citep{lay94,mundy96,lay97,kita02}.

\begin{figure}
\plotone{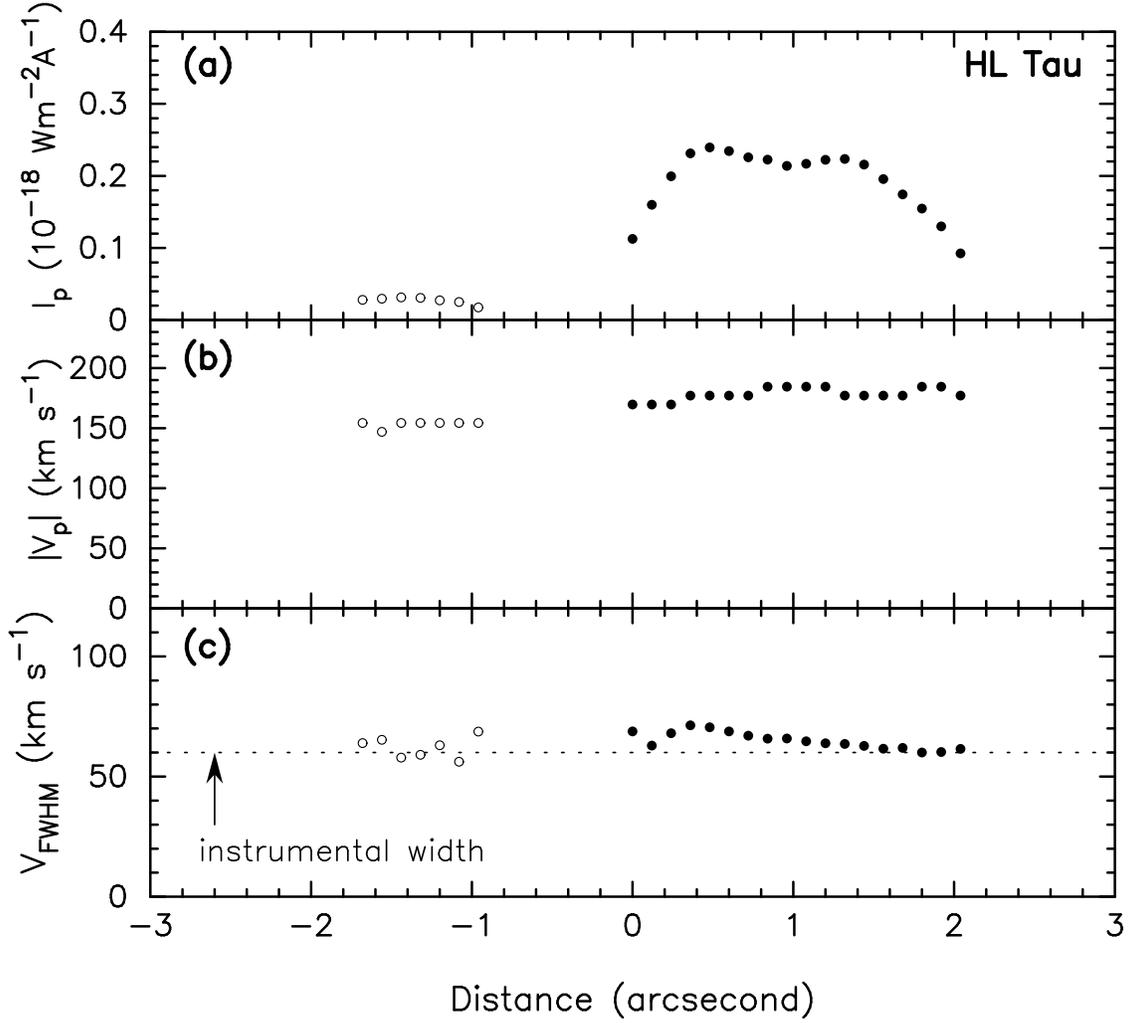}
\caption{($a$) The peak intensity (I$_p$), ($b$) absolute peak radial velocity ($|V_p|$), and ($c$) FWHM velocity width ($V_{\rm FWHM}$) of the \FeII\ emission are shown against the distance $Y$ from HL Tau.
The filled and opened circles trace the blueshifted and redshifted outflows, respectively.
The dotted horizontal line in $c$ represents the instrumental width of 60~$\kms$.
\label{fig3}}
\end{figure}

Figure~\ref{fig3} shows the variations of the peak intensity (I$_{\rm p}$), absolute peak radial velocity ($|{\rm V}_{\rm p}|$) with respect to the systemic velocity (V$_{\rm LSR}=+$6.4~$\kms$), and velocity width ($\Delta{\rm V}_{\rm FWHM}$) along the angular distance from HL~Tau at a 0$\rlap.^{''}$12 sampling.
The peak intensity I$_{\rm p}$ has two maxima in the blue-shifted jet ($Y\ge 0$): one at  $Y\sim $0$\dotsec$4 and the other at $Y\sim$1$\dotsec$4, as are also evident from Figure~1.

The blueshifted and redshifted jets have relatively constant peak velocities with their variations of the order of 5\%.
There is a noticeable asymmetry in the peak velocity between the two jets. 
The blueshifted jet has $|{\rm V}_{\rm p}| \sim$180~$\kms$, while the redshifted jet shows $|{\rm V}_{\rm p}| \sim$154~$\kms$.
This tendency is consistent with the \SII\ jet observed in 1987 February \citep{mundt90}, although the asymmetry was much larger in their paper.
The blueshifted \SII\ jet had the radial velocity of $-$200$\kms$ while the redshifted jet had $+$60 to $+$100$\kms$ within 10$''$ from the star when the measured heliocentric radial velocities are converted to the radial velocities with respect to the systemic velocity.

The observed velocity width of the \FeII\ jet shown in Figure~\ref{fig3}c is not much larger than the spectral resolution of 60~$\kms$, suggesting that the actual velocity width is narrower than $\Delta{\rm V}_{\rm FWHM}\sim$~40~$\kms$. 
The velocity width is largest at $Y\sim $~0$\dotsec$4 where the peak intensity has its first maximum. 
It then decreases monotonically with increasing distance from HL~Tau.
Such a decrease in the velocity width with distance might result from the progressive weakening of the LVC relative to the HVC.

\subsection{RW Aur A}

\begin{figure}
\plotone{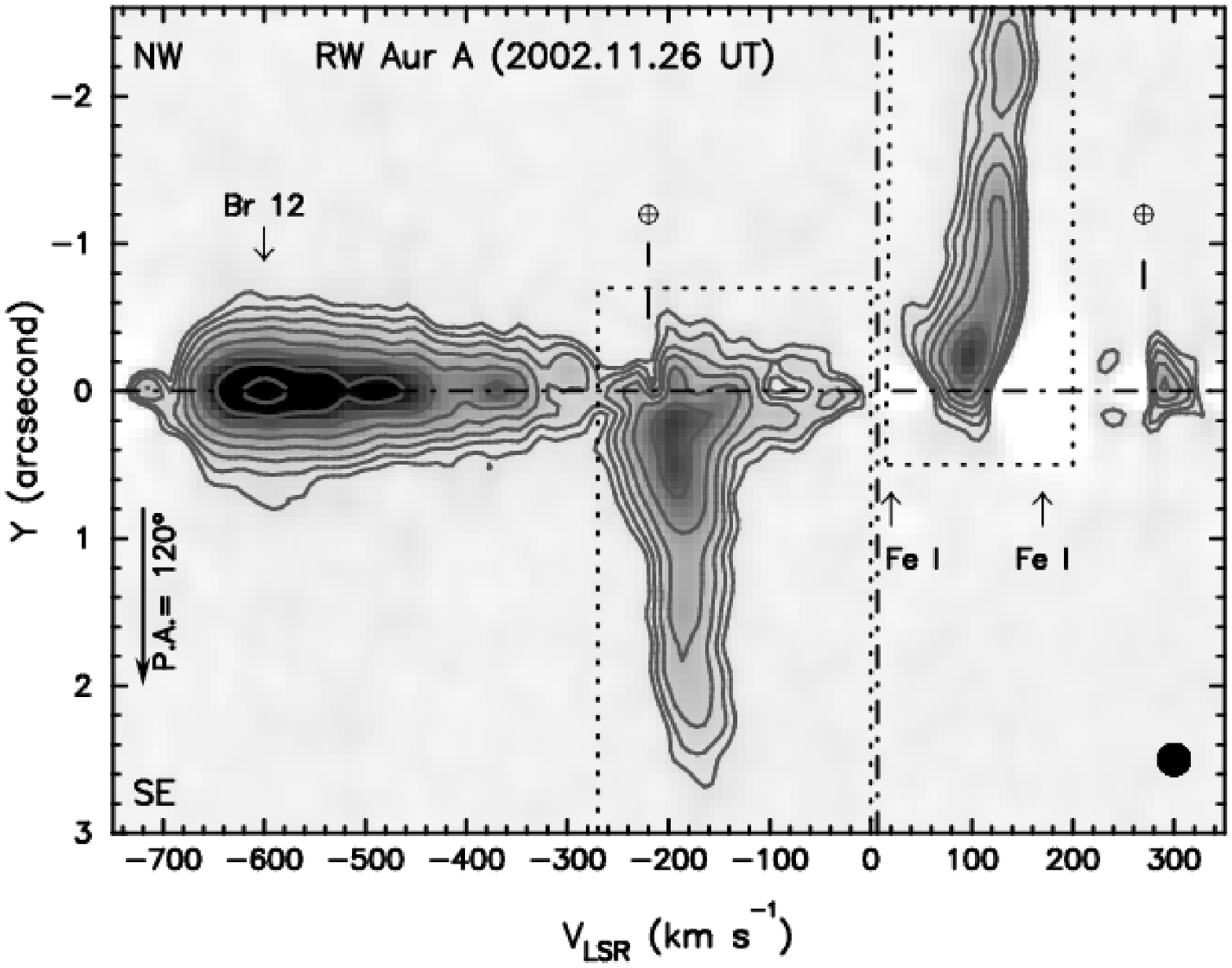}
\caption{Continuum-subtracted position velocity diagram of the \FeII\ $\lambda$1.644~$\micron$ emission toward the jet from RW~Aur.
The data was smoothed by a Gaussian with ${\sigma}_G~=$1.1 pixels. 
Contours are drawn from 0.03 ($\sim$3.6$\sigma$) to 0.73 ($\sim$88.0$\sigma$) $\times$ 10$^{-18}$ W m$^{-2}$ \AA$^{-1}$ with equal intervals in a logarithmic scale.
The axis $Y$ is along the slit with a position angle of 120$^\circ$.
The region with positive $Y$ corresponds to the southeast of RW~Aur.
The \FeII\ emission features are indicated by dotted line boxes. 
The dash-dotted vertical line indicates the systemic velocity at $V_{\rm LSR} =$ +6.0 $\kms$ \citep{UT87}.
The strong emission at $V_{\rm LSR} < -$270 $\kms$ is the Br12 emission.
Photospheric \ion{Fe}{1} lines are indicated by arrows and telluric lines are shown by $\oplus$. 
The filled ellipse at the lower right shows the velocity and spatial resolutions of 30 $\kms$ and 0$\rlap.^{''}$2, respectively.
\label{fig4}}
\end{figure}

The position-velocity diagram shown in Figure~\ref{fig4} is a continuum-subtracted \FeII\ $\lambda$1.644~$\micron$ emission spectrum along the slit, which was at the position-angle of 120$\degree$ and was unfortunately misoriented with respect to the actual position angle (130$^\circ$) of the RW~Aur jet \citep{hirth97, ME98, dougados00}. 
Both blueshifted and redshifted jets are seen together with a Br12 emission.
The blueshifted jet is seen at $Y > -$0$\dotsec$4 and $-$270~$< V_{\rm LSR}< -$10$~\kms$, where $Y$ is the projected angular distance measured along the slit from the star.
The redshifted counter jet is seen at $Y <$0$\dotsec$3 and 40~$< V_{\rm LSR}<$160~$\kms$.
Significant stellar continuum emission exists at $|Y| \lesssim~$0$\dotsec$1, where the PVD is affected by a weak periodic pattern caused by photospheric \citep[K4:][]{SBB00} and telluric absorption lines coupled with the detector undersampling effect \citep[see~][]{pyo03}.
A small dip around $Y = -$0$\dotsec$1 and $V_{\rm LSR}~\sim~-$90$~\kms$ was caused by this effect. 
The redshifted \FeII\ emission is peaked at $Y \sim -$0$\dotsec$2, with a marked drop toward the star apparently roughly consistent with the PSF size, suggesting that an optically thick disk around RW~Aur is smaller than 0$\dotsec$2 in the projected scale, or 40~AU in the actual scale. 
\citet{cabrit05} found a compact and bright CO structure centered on RW Aur A, suggesting a consistently small disk size of $\sim$40 AU. 

At the southeast of RW Aur ($Y > $0$\arcsec$), the entire \FeII\ emission is blueshifted with respect to the systemic radial velocity of $V_{\rm LSR} \sim +$6~$\kms$ \citep{UT87}. 
The blueshifted \FeII\ emission is dominated by a strong, narrow feature at $V_{\rm LSR} \sim -$180~$\kms$ extended along the jet.
As is similar to the case of HL~Tau, the narrow \FeII\ emission is accompanied by a less blueshifted wing or shoulder feature within 0$\dotsec$4 from the star.
This component covers the velocity from $-$10 to $-$100~$\kms$.
By analogy to the case of HL~Tau, we will call the more blueshifted, stronger and extended emission feature as the HVC and the less blueshifted wing or shoulder feature as the LVC.

A weak, low velocity emission component in the RW~Aur jet was also detected in the optical \SII\ and \OI\ lines \citep{hirth94,hirth97,HEG95}, while some observations showed negative results of such a component in the optical or infrared \citep{woitas02, davis03}.
The optical low velocity component has a peak velocity of 0--50~kms$^{-1}$ and was observed as part of the connecting emission between the blueshifted and redshifted jets at the stellar position.
The LVC of the \FeII\ line has only blueshifted emission and the emission does not connect the blueshifted and redshifted jets.
We must be careful, however, that the possible presence of photospheric \ion{Fe}{1} absorption at $+$20~kms$^{-1}$ could eliminate otherwise detectable emission around this velocity.

\begin{figure}
\plotone{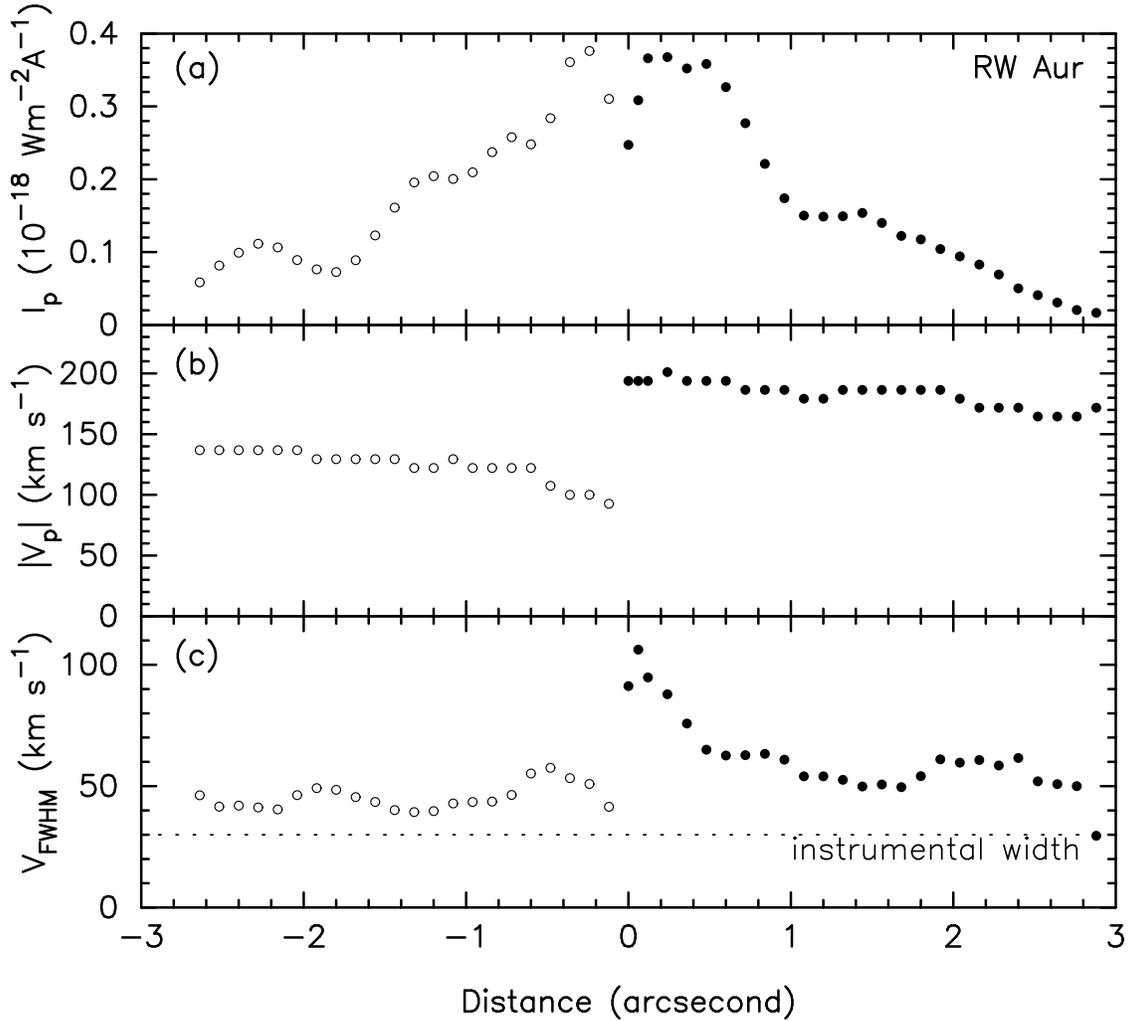}
\caption{($a$) The peak intensity (I$_p$), ($b$) absolute peak radial velocity ($|V_p|$), and ($c$) FWHM velocity width ($V_{\rm FWHM}$) of the \FeII\ emission are shown against the distance $Y$ from RW~Aur.
The filled and opened circles trace the blueshifted and redshifted outflows, respectively.
The dotted horizontal line in $c$ represents the instrumental width of 30~$\kms$.
\label{fig5}}
\end{figure}

Figure~\ref{fig5} shows the variations of the peak intensity (I$_{\rm p}$), absolute peak radial velocity ($|{\rm V}_{\rm p}|$) with respect to the systemic velocity (V$_{\rm LSR}=+$6~$\kms$), and velocity width ($\Delta{\rm V}_{\rm FWHM}$) along the angular distance from RW~Aur at a 0$\rlap.^{''}$12 sampling.
The peak intensity decreases with distance from the star, but this is largely due to the misoriented slit with respect to the actual position angle of the jet (130$\degree$).
The jet is well-collimated with the FWHM width smaller than 0$\dotsec$4 within 3$\arcsec$ \citep{dougados00,woitas02}.
Thus the slit of 0$\dotsec$3 width misoriented by 10$^\circ$ with respect to the jet will completely depart from it at $|Y|\sim$~3$\arcsec$ if we assume that the jet has a FWZI width of 0$\dotsec$8.

The decrease of I$_{\rm p}$ toward the stellar position at $|Y|\la$~0$\dotsec$1 has an important meaning.
It is associated with the fact that the HVC is extended and located further away from the star than the LVC, as has been observed in the near-infrared \FeII\ emission for other objects including HL~Tau in the current study \citep{pyo02, pyo03}.
Decrease in forbidden line intensities toward jet driving stars was also observed in the \OI\ and \SII\ emissions and has been interpreted as a result of rapid collisional damping due to a higher density in the vicinity of the stars \citep{hirth97, woitas02}.

The blueshifted and redshifted jets have similar peak intensities in the current \FeII\ data, while the redshifted \FeII\ emission was stronger than the blueshifted one in 2001 November \citep{davis03}.
It is difficult to investigate the cause of this because the angular resolutions of the two observations are largely different.  
In addition, the slit we used had a narrower width and was misoriented with respect to the actual jet axis.  
Redshifted emissions of optical forbidden lines (\SII, \OI\ and \NII) within 10$^{''}$ from the star were also stronger than the corresponding blueshifted emissions between 1993 to 2000 \citep{woitas02, dougados00, ME98, hirth97, HEG95, hirth94, hamann94}.

Figure~\ref{fig5}$b$ shows that the jet velocities have large asymmetry with respect to the systemic velocity.
The blueshifted jet has $|{\rm V}_{\rm p}|=$170 -- 200~$\kms$, while the redshifted jet shows $|{\rm V}_{\rm p}|=$100 -- 140~$\kms$.
The velocity asymmetry in the RW~Aur jet has been well known since it was discovered by \citet{hirth94} with optical long-slit spectroscopy.
Although the asymmetry seems to exist even in the close vicinity of the star, we must be careful that the presence of a photospheric \ion{Fe}{1} absorption line may have affected the peak velocity of the redshifted jet at $|Y|\la$~0$\dotsec$4.
The \ion{Fe}{1} line is located at V$_{\rm LSR}=$~170~kms$^{-1}$ with respect to the rest wavelength of the \FeII\ line as indicated in Figure~3.
We can thus safely say that the origin of velocity asymmetry is located within 0$\dotsec$4 from the star.

The absolute velocity decreases with distance from the star for the blueshifted jet, while it increases for the redshifted jet.
This tendency was also reported by \citet{davis03}.
It is impossible to produce such an asymmetry by any symmetric velocity structure between the blueshifted and redshifted jets combined with the misoriented slit effect.
Thus the velocity asymmetry itself may be time variable.

Figure~\ref{fig5}$c$ shows the spatial variation of the velocity width. 
In the blueshifted jet, the width is $\sim$100~$\kms$ near the star (0$'' < Y \la 0\dotsec2$) and decreases to 50 -- 60~$\kms$ for $Y \ga$~0$\dotsec$5.
In the redshifted jet, the width is $\sim$50~$\kms$ and shows no evidence of increase near the star.
In fact, the velocity width decreases toward the star ($-$0$\dotsec$4$ \la Y <$0$''$). 
This may be caused by the presence of the photospheric \ion{Fe}{1} absorption line at V$_{\rm LSR}=$170~$\kms$.

\section{DISCUSSION}

\subsection{The Emission Source Locations around HL Tau}

The LVC of HL~Tau seems to be strongest at the stellar position with a weaker emission significantly extended to the northeast ($Y >$~0, see Figure~\ref{fig2}).
Because HL~Tau has a scattering envelope extended to the northeast \citep{close97}, we must first examine whether the extended LVC emission comes from scattering.

\begin{figure}
\plotone{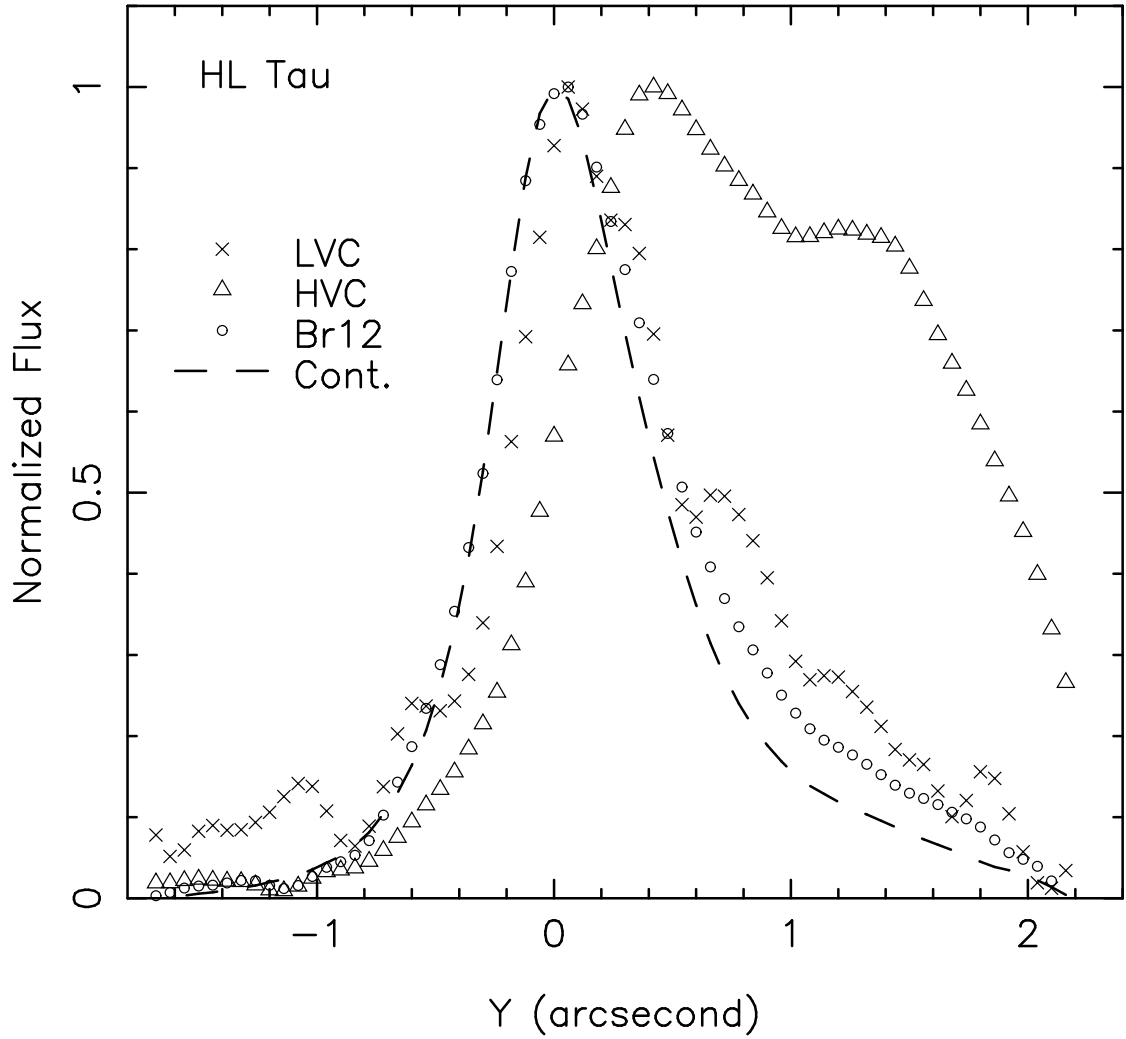}
\caption{Normalized spatial intensity profiles of the low velocity component ($\times$), high velocity component ($\triangle$), Br12 emission ($\circ$) and continuum (dashed line) for HL~Tau.
\label{fig6}}
\end{figure}

Figure~\ref{fig6} shows the normalized spatial profiles of intensity for the LVC, HVC, Br12 and continuum emissions. 
The intensities for the LVC and HVC are averaged over the velocity ranges of $-$100 to 0~$\kms$ and $-$270 to $-$100~$\kms$, respectively.
The Br12 intensity is averaged over $-$700 to $-$280~$\kms$ with respect to the rest wavelength of \FeII\ \footnote{$-$190 to $+$231 $\kms$ with respect to the rest velocity of Br12 ($\lambda$1.6412 $\mu$m)}, and the continuum emission is averaged over $+$400 to $+$800~$\kms$ and $-$1200 to $-$1000~$\kms$ where the spectrum is free from photospheric absorption lines.

The continuum, Br12 and LVC emissions show similarly skewed spatial profiles with sharp cutoffs on the redshifted side ($Y <$~0$''$) and gradual decreases on the blueshifted side ($Y >$~0$''$).
The normalized spatial profile of Br12 exactly matches that of the continuum on the redshifted side ($Y <$~0$''$), reaching its half maximum at $Y = -$0$\dotsec$28 consistent with the spatial resolution.
This means that the dominant source of the Br12 emission coincides with that of the continuum, namely the star itself or its unresolved vicinity.
We should note, however, that the peak position of the Br12 spatial profile is displaced by 0$\dotsec$038 $\pm$ 0$\dotsec$02 with respect to that of the continuum in Figure~5.
The small but finite amount of the observed spatial displacement between the Br12 and continuum peaks may mean that at least part of the Br12 emission at $Y\sim$ 0$''$ arises in the unresolved outflow.
This conclusion is consistent with the recent observations suggesting that infrared \ion{H}{1} lines of classical T~Tauri stars or Class~I objects originate in the outflow acceleration region \citep{FE01, NAG04, WRD04}, although existing outflow models as well as magnetospheric accretion models still have difficulties to quantitatively explain their line profile characteristics.

On the blueshifted side ($Y >$~0$''$), the gradual decrease of the continuum emission is caused by the scattering of the stellar light off the cavity located at the northeast of HL~Tau \citep{close97}.
Thus the spatial profiles of Br12 and LVC similar to that of the continuum suggest that their emissions in the northeast of HL~Tau arise mostly from scattering.
However, the normalized spatial profile of Br12 is significantly larger than that of the continuum on the blueshifted side ($Y >$~0$''$).
The Br12 emission along the blueshifted jet, therefore, does not entirely arise from scattering of the central unresolved light source, but part of it must originate in the extended blueshifted jet itself.

The normalized spatial profile of LVC is more interesting.
Although it basically follows a skewed shape similar to those of Br12 and the continuum emission, the spatial profile of LVC in the southwest ($Y <$~0$''$) is offset by $\Delta Y = +$0$\dotsec$077 $\pm$ 0$\dotsec$04 with respect to that of the continuum.
This means that the dominant emission of LVC arises not from the unresolved vicinity of the star but in the jet $\sim$~15 AU away from the star.
In the northeast ($Y >$~0$''$), the normalized intensity of LVC is larger than that of the continuum, although the excess is not much larger than the noise level of the LVC profile.
Part of the extended LVC emission in the northeast thus also arises in the jet.

Figure~\ref{fig4} shows that the LVC of RW~Aur also has its spatial maximum displaced to the blueshifted (southeast) side. 
However, this may be caused by the presence of the artificial dip around $Y=~-$0$\dotsec$1 and V$_{\rm LSR}=$~90~km~s$^{-1}$ and is not reliable.

\subsection{The High and Low Velocity Components} \label{diss_2v}

Our observations revealed that the \FeII\ emission lines show  both LVC and HVC in HL~Tau and RW~Aur.
The presence and nature of the two velocity components are similar to the cases of L1551 IRS~5 and DG~Tau \citep{pyo02, pyo03}. 
The LVCs of HL~Tau and RW~Aur are, however, not as distinct as those of L1551 IRS~5 and DG~Tau seen in their PVDs.
The latter sources show that the LVC has a clearly separated peak located away from the stellar position in the PVD, while the former sources have the LVC observed only as a wing or shoulder of the HVC and its spatial location is close to, although not necessarily coincident with, the stellar position.
For the jets from L1551 IRS~5 and DG~Tau, we suggested that the HVC is physically different from the LVC because of the largely different characteristics of the two components. 
In fact, it has not been easy to model the two distinct components with single outflow models \citep{pesenti03, shang98}.
In the cases of HL~Tau and RW~Aur, however, the LVC seems to be more physically related to the HVC from their appearance in the PVDs, and we will re-examine the relation between the two velocity components for these objects below\footnote{Please refer to \citet{pyo02,pyo03} for more details about two velocity components.}.

Up to date the magnetocentrifugal force is widely considered as the most comprehensive mechanism for launching outflows from the star-disk systems of young stellar objects \citep[see reviews in][]{KP00,Shu00}.
There are two kinds of popular models based on the magnetocentrifugal mechanism: the {\it disk wind model} \citep{KP00} and {\it X-wind model} \citep{Shu00}.
The disk wind model assumes that the outflow launching region is located over a wider range of disk radii outside the inner disk edge, while the X-wind model postulates that the launching region is located at a very narrow disk radii close to the inner disk edge where the highly concentrated stellar magnetic field interacts with the disk.
Synthetic PVDs of the forbidden emission lines predicted by their two models may be compared with observations \citep[e.g.][]{mtakami06}.

\begin{figure}
\epsscale{0.6}
\plotone{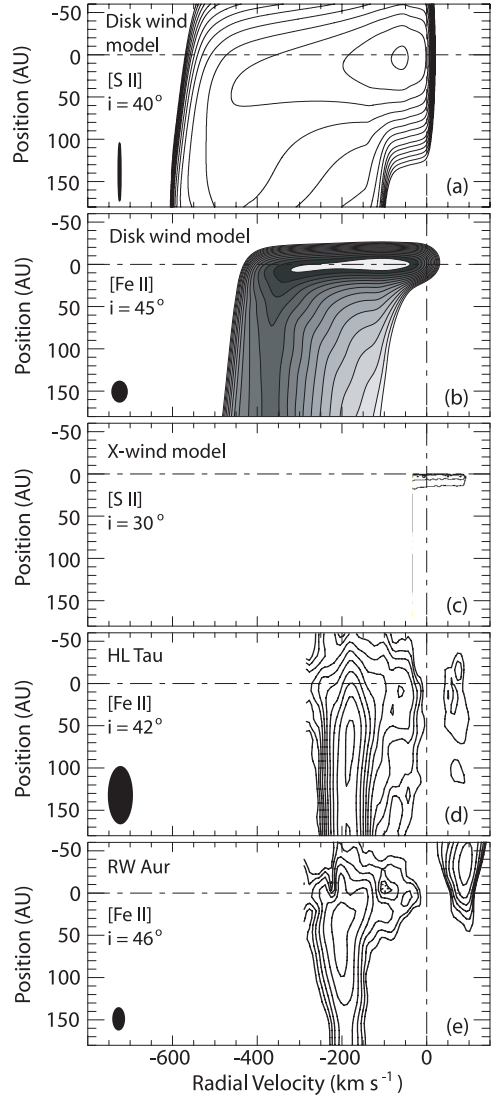}
\caption{Comparison of model PVDs ($a$, $b$ and $c$) with the present observations ($d$ and $e$).
$i$ is the inclination angle of the jet axis with respect to the line of sight.
The filled ellipses at the lower left corners show the velocity and spatial resolutions.
($a$) Cold disk wind Model 1 by \citet{cabrit99} for \SII~$\lambda$6731 \AA\ and $i =$ 40$\degree$.  
Convolved with a beam of 70 AU $\times$ 10 $\kms$.
($b$) Cold disk wind Model A by \citet{pesenti03} for \FeII~$\lambda$1.644$\mu$m and $i =$ 45$\degree$. 
Convolved with a beam of 22.4 AU $\times$ 30 $\kms$.
($c$) X-wind model by \citet{shang98} for \SII~$\lambda$6731 \AA\ and $i =$ 30$\degree$.
($d$) Present data for HL~Tau. 
$i =$ 42$\degree\pm$5$\degree$ and the resolution is 70 AU $\times$ 60 $\kms$.
Note that most of the low velocity emission at $V > -$100 $\kms$ and position $>$ 50 AU is caused by scattering.
($e$) Present data for RW~Aur.
$i =$ 46$\degree \pm$3$\degree$ and the resolution is 28 AU $\times$ 30 $\kms$.
\label{fig7}}
\end{figure}

In Figure~\ref{fig7} we compare the PVDs of HL~Tau and RW~Aur with those of {\it the cold disk wind models} \citep{cabrit99, garcia01, pesenti03} and {\it X-wind model} \citep{shang98}.
The two models show a similitude that their velocity widths are broad at the stellar position and become narrower when the position moves away from the star.
The broad velocity width at the star is caused by the presence of lower velocity emission added to the radial velocity V $\la$ 100 $\kms$, while its spatial extent (20 -- 100 AU) depends on the adopted spatial resolutions of the models.
This emission arises in the region where uncollimated outer stream lines have sufficient emissivity in these models.

The above noted characteristic of the model PVDs are also apparent in the observed PVDs of HL~Tau and RW~Aur.
By analogy to the observed PVDs, we call the lower velocity emission at V $\la$ 100 $\kms$ the LVC and the spatially extended higher velocity emission at V $\ga$ 100 $\kms$ the HVC.
The observed PVDs are, however, different from the model PVDs on the following points.
\begin{enumerate}
\item{The HVC of the {\it disk wind models} shown in Figure~\ref{fig7}$a$ and \ref{fig7}$b$ shows a velocity width of 200 -- 500 $\kms$, much larger than the observed width of $\la$ 60 $\kms$.
On the other hand, the width of HVC for the {\it X-wind model} shown in Figure~\ref{fig7}$c$ is narrow, which may be consistent with the observed width of HVC and its symmetrical line profiles if internal velocity fluctuation in the HVC is taken into account.
Recently \citet{pesenti04} showed that a ``warm'' disk model can make narrower line widths, showing much better agreement with those of observed jets.}
\item{The LVC of the {\it X-wind model} shows a significant redshifted emission arising from the streamlines on the far side.
The presence of such redshifted emission coincident with the star is marginal in the observed PVDs because of the contamination by the photospheric \ion{Fe}{1} absorption at $+$20 $\kms$ (see \S3.1).
On the other hand, the LVC for the {\it disk wind models} has little redshifted emission at the stellar position, which is consistent with the observed characteristics of LVC unless the marginal emission is real.}
\item{The HVC of the {\it cold disk wind models} exhibits significantly higher peak velocity by a factor of 2--2.5 than the observed HVC does, while the HVC peak velocity of the {\it X-wind model} is quantitatively consistent with the present observations.
The {\it disk wind models} can be accommodated to the observations if a smaller Alfv\'en radius by a factor of 2-2.5 is adopted.
\citet{garcia01} also indicated that a denser and slower ``warm'' disk wind are favored to solve this problem.}
\end{enumerate}

With respect to the first point, material is launched from a large range of disk radii (0.07 -- 2 AU) in the disk wind models, where lower velocity material is ejected from the outer radii in the launch zone, while higher velocity material is ejected from the inner radii (the terminal velocity scales as the Keplerian speed at the jet base). 
The broad velocity width of HVC is hence caused by the outer streamlines having significantly lower radial velocities than the inner streamlines even when they are well collimated.
On the other hand, all the streamlines of the X-wind have similar Alfv\'en radii because they originate from the same launching point at the inner disk edge ($\sim$ 0.07 AU), resulting in similar radial velocities when they are collimated. 
The narrow velocity width of the observed HVC thus means for the disk wind models that the outer streamlines with lower velocities are associated with little emission, with a rapid cutoff of emission from the inner to outer streamlines, or that all the streamlines originate from a narrow radial range, which is rather similar to the case of an X-wind.
The presence of weak but spatially extended emission in the observed LVC, however, suggests that the extended LVC may correspond to the emission from the outer, low velocity streamlines that originate at larger radii than those of HVC. 

Regarding the second point, the X-wind model has the streamlines at the base of the jet fanning out in a wide direction, toward and away from the observers. 
As a result, a very broad pedestal emission of $\sim$ 300 $\kms$ wide is predicted within 20 AU from the star at all inclinations, extending to $+$100 $\kms$ on the redshifted side.
The X-wind model may be accommodated to the observed PVDs if the outer streamlines with higher redshifted and blueshifted radial velocities give little emission, unless the marginal redshifted emission observed toward HL~Tau is real.  
Another possibility is that the redshifted emission is weak or hidden in the optical wavelength where the extinction across the outflow tends to be large near the disk surface.

In summary, none of the published model PVDs are fully consistent with the observed PVDs of \FeII\ emission from HL~Tau and RW~Aur.\footnote{It is worth noting that the PVDs of \SII\ may differ from those of \FeII\ because the former line has a lower critical density than the latter one.}  
The narrow velocity width and symmetrical line profiles of the HVC seem to be more consistent with the X-wind model than the disk wind model, while the observed LVC emission located away from the star is in favor of the disk wind model.
Given the difficulty to consistently explain the observed PVDs with a single model, we may need a hybrid type model, e.g. an X-wind type outflow emanating from the truncation radius combined with a disk wind in the outer region \citep[see e.g.][]{pyo03}, although \citet{FDC06} discussed recently that it may be difficult for an X-wind to co-exist with an extended disk wind.

\subsection{The Truncation Radius of the HL Tau Disk}

For HL~Tau the peak intensity of the redshifted jet is 10 times smaller than that of the blueshifted jet when measured at the same projected distance of 1\rlap.{$''$}4 or 200~AU from the star.
Assuming the same intrinsic intensities for the blueshifted and redshifted jets, we estimate that the redshifted jet has an excess extinction of $\Delta A_{H} = $2.5~mag, corresponding to $\Delta A_{V} = $14~mag, with respect to the blueshifted jet.
This means that a significant amount of envelope material is present around HL~Tau, as is consistent with its nature as being still embedded in a molecular cloud.
In fact the optical \SII\ emission shows a larger gap of $\sim$3$\arcsec$ between the redshifted outflow and the star \citep{solf89}. 
The optical gap size is similar to the size of the northeast scattering envelope observed by \citet{Sta95} and \citet{close97}, suggesting that the extinction is caused in the relatively spherical envelope surrounding HL~Tau and its disk.

The visual extinction of $A_{V}=$14~mag at the projected radius of 200~AU corresponds to the surface mass density of 0.05~g~cm$^{-2}$ or the average density of 2$\times$10$^{6}$~cm$^{-3}$ if the excess extinction occurs along the line of sight length of 400~AU near the star.
The surface mass density can be compared to that of the disk at the same projected radius.
The disk has a surface mass density of 2 -- 4~g~cm$^{-2}$ at a radius of 100~AU \citep{kita02}, giving an expected surface mass density of 0.4 -- 0.8~g~cm$^{-2}$ corrected for inclination at the projected radius of 200~AU when we follow the two models described by \citet{kita02}.
The expected surface density of the disk is still one order of magnitude larger than that derived from the excess extinction, suggesting that the disk should be truncated abruptly at some projected radius smaller than 200~AU from the star, most probably near the projected radius of 0$\dotsec$8, which is the gap size between the redshifted outflow ($>$ 3$\sigma$) and central star, corresponding to the actual radius of 160~AU.

\subsection{Knots in the RW Aur Jet}
The jet from RW~Aur has knots in optical forbidden line emissions \citep{dougados00, woitas02, LCD03}, with the redshifted jet showing more knots than the blueshifted one. 
The current \FeII\ data also shows a hint of knots along the jets.
In Figure~\ref{fig5}$a$, the redshifted jet intensity shows four local maxima at $Y =$ $-$0$\dotsec$25, $-$0$\dotsec$73, $-$1$\dotsec$20, and $-$2$\dotsec$29, and the blueshifted jet intensity has two local maxima and one shoulder at $Y =$ 0$\dotsec$17, 0$\dotsec$48, and 1$\dotsec$25, respectively.
We compared these positions with the results by \citet{LCD03}, who presented the proper motions of knots within $\sim$12$^{''}$ from the star observed in \SII\ and \OI\ lines from 1997 to 2000.
The results of the comparison are summarized as follows:

\begin{enumerate}

\item{
The local maxima at $Y =$ $-$0$\dotsec$25 and 0$\dotsec$17 may be stationary because knots were observed at similar positions in 1998 and 2000 in the optical forbidden lines.  }
\item{
The local maxima at $Y =$ $-$1$\dotsec$20 and $-$2$\dotsec$29 correspond to the knots R6 and R5, respectively, of \citet{LCD03} when their proper motions of 0$\dotsec$16~yr$^{-1}$ and 0$\dotsec$24~yr$^{-1}$, respectively, are taken into account.}
\item{
The shoulder around $Y =$ 1$\dotsec$25 $\pm$ 0$\dotsec$15 may correspond to the knot B4 if it has a proper motion of 0$\dotsec$34$\pm$0$\dotsec$1~yr$^{-1}$.}
\item{
The local maxima at $Y =$ $-$0$\dotsec$73 and 0$\dotsec$48 may be newly ejected knots because they are close to the source and were not seen in \citet{LCD03}.}   
\end{enumerate}   

In general, the local maxima observed in the \FeII\ intensity spatial profile show good agreement with the optical forbidden line knot positions if their proper motions are taken into consideration.
Because \FeII\ has a larger critical density than [\ion{S}{2}] \citep{hamann94b,reipurth00}, \FeII\ traces denser regions of knots detected in [\ion{S}{2}].
 
\section{SUMMARY}
We observed the jets from HL~Tau and RW~Aur~A in the \FeII\ $\lambda$~1.644~$\mu$m emission line with high angular and spectral resolutions using the Subaru Telescope adaptive optics system.
The main results are summarized as follows:

\begin{enumerate}
\item{
The blueshifted jet for each object shows two emission features: a strong and spatially extended emission with a peak velocity of $-$200 to $-$170~$\kms$ and a less blueshifted wing or shoulder emission covering $-$100 to 0~$\kms$ localized to the stellar position.
By analogy with the \FeII\ jets from L1551 IRS~5 and DG~Tau, we call them the high (HVC) and low (LVC) velocity components, respectively.
}

\item{
The position velocity diagrams (PVDs) for the blueshifted jets from HL~Tau and RW~Aur are similar to those of L1551 IRS~5 and DG~Tau except that the LVCs for the former objects are not clearly detached from the stellar positions.
Because of this, they are similar to the synthesized PVDs of single jet model calculations.
}

\item{
For HL~Tau, however, detailed comparison between the spatial profiles of the LVC and continuum revealed that peak of the LVC is displaced by $\sim$0$\dotsec$077 $\pm$ 0$\dotsec$04 from the star and is extended $\sim$1$\arcsec$ along the jet, suggesting that peak of the LVC emission does not coincide with the stellar position but is located $\sim$15~AU away from the star and is extended over $\sim$200 AU.
}

\item{
The observed PVDs are compared with the synthesized PVDs of cold disk wind and X-wind models.
The disk wind model tends to have too large line width at the jet, while the X-wind model has excess redshifted emission at the stellar position.
The narrow velocity width of the observed high velocity emission supports an X-wind type model where the launching region is localized in a small radial range, while the low velocity emission located away from the star favors the presence of a cold disk wind.
}

\item{
For HL Tau, the normalized spatial profile of Br12 is significantly extended compared with the normalized continuum profile along the blueshifted jet.
It indicates that the Br12 emission originates not only from the central unresolved region but also in the blueshifted jet itself.
}

\item{
The redshifted counter jets were detected from both HL~Tau and RW~Aur. 
The PVD of HL~Tau shows a gap of 0$\dotsec$8 between the redshifted jet and the star, which is caused by the presence of an optically thick circumstellar disk of $\sim$160 AU in radius.
The redshifted jet from RW~Aur~A shows a sharp drop toward the star from its peak located at Y$\sim -$0$\dotsec$2, suggesting that its disk radius is smaller than 40~AU.
}

\item{
The local intensity maxima in the RW~Aur jet show good agreement with the knot positions identified in optical forbidden line emissions when their proper motions are taken into account.
}

\end{enumerate}

\acknowledgments
We thank the referee for useful comments and thoughtful suggestions.
We are grateful to the entire staff of the Subaru Telescope for their dedicated support to the telescope and observatory operations. 
This research has made use of NASA's Astrophysics Data System and the SIMBAD database which is operated at CDS, Strasbourg, France.


\clearpage

\clearpage

\clearpage

\clearpage

\clearpage

\clearpage

\clearpage

\end{document}